\documentstyle[twocolumn,epsfig]{mn}
\begin{document}

\title{Properties of vertically self-gravitating accretion discs with a dissipative corona}

\author[F. Khajenabi $\&$ P. Duffy]{Fazeleh Khajenabi$^{1}$\thanks{E-mail:
fazeleh.khajenabi@ucd.ie (FK); } and {Peter Duffy$^{1}$\thanks{E-mail: peter.duffy@ucd.ie (PD)}}\\
$^{1}$School of Physics, University College Dublin, Belfield, Dublin 4, Ireland}

\maketitle

\date{Received ______________ / Accepted _________________ }

\begin{abstract}
The steady-state structure of a disc with a corona is analyzed when the vertical component of the gravitational force due to the self-gravity of the disc is considered. For the energy exchange  between the disc and the
corona, we assume a fraction $f$ of the dissipated energy inside the accretion disc is transported to the corona via the magnetic tubes. Analytical solutions corresponding to a prescription for $f$ (in which this parameter directly depends on the ratio of the gas pressure to the total pressure) or free $f$ are presented and their physical properties are studied in detail.
We show that the
existence of the corona not only decreases the temperature of the disc, but also increases
the surface density.The vertical component of the gravitational force due to the self-gravity of the disc decreases the self-gravitating radius and the mass of the fragments at this radius. However, as more energy is transported from the disc to the corona, the effect of the vertical component of the gravitational force due to the self-gravity of the disc on the self-gravitating radius becomes weaker, though the mass of the fragments is reduced irrespective of the amount of the energy exchange from the disc to the corona.

\end{abstract}

\section{Introduction}
Accretion plays a crucial role in the processes of energy liberation and mass accumulation near
the cores of compact astronomical
objects (Bertout  1989;  Robinson  1976; Hunter, Ball  \&  Gottesmanet  1984; Kohler 1995), and
in particular for Active Galactic Nuclei (AGN) (Shields 1978 ;Osterbrock 1993;Szuszkiewicz 2001;
Czerny 2007). Self-gravity has global consequences on the disc shape, influencing the
location  of its inner and outer edges, as well as the disc geometrical thickness
(Karas, Hur\'e \& Semer\'ak 2004). The role of self-gravity has been separated into that from
its vertical component (e.g., Fukue \& Sakamoto 1992;
Hur\'e 1998) or from its radial component (Lu, Yang \& Wu 1997), and from combination of two (Yang \& Liu 1990;
Bertin \& Lodato 1999). The vertical structure of accretion discs of galactic nuclei is affected
by their own self-gravity and the gravitational force of the central object, especially in the middle
and outer regions, where the vertical component of the disc gravitational field exceeds the
corresponding vertical component due to the central object.

In addition to the role of self-gravity, one of the main open issues
in the physics of black hole accretion discs is the relationship
between the disc MRI-driven turbulent viscosity and the generation
of a hot corona, which is postulated in order to explain the
observed X-ray emission (Liang \& Price 1977; Galeev, Rosner \&
Vaiana 1979; Blackman \& Field 2000; Kuncic \& Bicknell 2004).
Furthermore, theoretical simulations successfully reproduce the
observed spectra by assuming a certain fraction of gravitational
energy is released as hot gas (e.g., Haardt\& Maraschi 1991; Svensson \& Zdziarski 1994; Kawaguchi,
Shimura \& Minishige 2001) and/or a certain spatial distribution of
the hot gas (e.g., Merloni \& Fabian 2001b). Svensson \& Zdziarski (1994) studied a disc-corona system, in which a major fraction of the power released when the cold matter accretes is transported to and dissipated in the corona.  They showed that the steady-state structure of the disc significantly modified due to the existence of a corona. Miller \& Stone (2000)
showed that the primary saturation mechanism of the MRI is local
dissipation, about twenty five percent of magnetic energy generated
by MRI within two scale heights escapes because of buoyancy,
producing a  strongly magnetized corona above the disc. It was shown
by Merloni \& Fabian (2001a) that the inferred thermal energy
content of the corona, in all black hole systems, is far too low to
explain their observed hard X-ray luminosities, unless either the
size of the corona is at least of the order of $10^3$ Schwarzschild
radii, or the corona itself is in fact a reservoir, where the energy
is mainly stored in the form of a magnetic field generated by a
sheared rotator.

Following these considerations Khajenabi \& Shadmehri (2007) (hereafter KS) studied the gravitational instability of an accretion disc
with a corona. They showed that the disc becomes more
gravitationally unstable due to the existence of the corona. Because
of the corona, not only does the self-gravitating radius decrease,
but also the mass of the fragments at this radius increases (KS).
However, KS did not take into account the vertical component of the
gravitational force due to  the self-gravity of the disc. Also,
their analysis was based just on the Toomre parameter criteria.
Actually, theory and simulations (Gammie 2001; Rice et al. 2003,
2005; Lodato \& Rice 2004; Mejia et al. 2005) show that {\it two}
conditions should be fulfilled for accretion discs to fragment
gravitationally. First, the accretion disc must be massive enough so
that gravity can overcome thermal pressure and centrifugal support,
which implies a value of less than unity for the Toomre parameter
(Toomre 1964). Second,  the disc must cool fast enough for the
compressional energy provided by the collapse to be radiated away.
So, according to the numerical simulations (Gammie 2001; Rice et al.
2003, 2005; Lodato \& Rice 2004; Mejia et al. 2005), the cooling
time-scale must be of the order of the dynamical time-scale. For an
accretion disc with a corona, we believed that the corona provides another
mechanism for the cooling of the disc and so, existence of the
corona decreases the cooling time-scale.

In this paper, we extend the analysis of KS in different directions:
(i) Toomre parameter  and the cooling time-scale conditions are
considered as criteria for the fragmentation of the disc; (ii) the
vertical component of the gravitational force due to the
self-gravity of the disc is considered; (iii) for the fraction of
the dissipated energy inside the disc that is transported  to the
corona, we consider two cases. in the first case, this fraction
depends on the physical variables of the system (Merloni \&
Nayakshin 2006); and in the other case, this fraction
is assumed to be constant and we consider it as a free parameter of
the model. In the next section, the basic assumptions and the
equations are presented. We analyse properties of the
solutions in Section 3. The gravitational stability of our solutions
is studied in detail in Section 4 .  By studying the gravitational
stability of such a system, we address the effect of self-gravity on
disc stability and fragmentation. We conclude with a summary of the
results and a discussion in the  final section.

\section{General formulation}
As a result of differential rotation the gas is heated, loses angular momentum and a slow (subsonic)
radial drift results. In a steady state the mass accretion rate is  constant,
and the local viscous energy release is balanced by the radiative cooling. In the direction
perpendicular to the plane of the disc, hydrostatic equilibrium is assumed.
The rotation curve is dominated by a Newtonian point mass $M$, as
relativistic effects are only important at small radii. Thus, the rotational angular velocity of the
disc is approximately Keplerian, i.e. $\Omega_{\rm K}=\sqrt{GM/R^{3}}$. Although we neglect the radial component of
the self-gravity of the disc,the self-gravity of the disc in the vertical direction that modifies  the
hydrostatic equilibrium of the disc is  included in our model. We take it into account with the
extra assumption that the accretion disc is homogeneous and extends to infinity, as done in several other
models (Paczynski 1978; Kozlowski, Wiita \& Paczynski 1979; Sakimoto \& Coroniti 1981; Shore \& White 1982;
Cannizzo \& Reiff 1992; Hure et al. 1994a). The weight of the corona is negligible as has been shown by Svensson \& Zdziarski  (1994). So, the vertical hydrostatic equilibrium of the disc implies

\begin{equation}\label{eq: Zdirec}
\frac{p}{\Sigma}=\frac{\Omega_{\rm K}^2 H}{2}(1 +\frac{4 \pi G \rho}{\Omega_{\rm K}^2 }),
\end{equation}
where we introduce
\begin{equation}
\zeta =\frac{4 \pi G \rho}{\Omega_{\rm K}^2 },\label{eq:zetaf}
\end{equation}
which measures, along the axis of rotation, the acceleration due to self-gravity as a fraction of the
central gravitational acceleration. With this definition, equation (\ref{eq: Zdirec}) can be rewritten as

\begin{equation}
\frac{p}{\Sigma}=\frac{\Omega_{\rm K}^2 H}{2}(1 +\zeta).\label{eq:zcom}
\end{equation}

In what follows, when we discuss  solutions without self-gravity of the disc, we mean solutions that are obtained by neglecting $\zeta$ in the above equation (see KS). Also, by solutions with self-gravity of the disc, we mean solutions that are obtained by considering $\zeta$ in the above equation. We obtain solutions with self-gravity in this paper and illustrate the possible effect of the self-gravity by making comparisons to solutions without self-gravity.

The basic remaining equations of our model are similar to KS. However, we briefly mention the equations and
basic assumptions for completeness. We  consider a more general prescription for the viscous stresses
$\tau_{\rm r\phi}$ (Taam \& Lin 1984; Watarai \& Mineshige 2003; Merloni \&  Nayakshin 2006):
\begin{equation}
\tau_{\rm r\phi}=-\alpha_{0} p^{1-\mu/2} p_{\rm gas}^{\mu/2},\label{eq:visg}
\end{equation}
where $\alpha_{0}$ and $0\leq\mu\leq2$ are constants and $p$ is
the sum of the gas and radiation pressures. Phenomenological  models
generally assume that at each radius, a fraction $f$ of the accretion
energy is released in the reconnected magnetic corona. Assuming that in
MRI-turbulence discs such a fraction $f$ of the binding energy is
transported from large to small depths by Poynting flux, Merloni \& Nayakshin
(2006) estimated the fraction $f$ as
\begin{equation}
f=\sqrt{2\alpha_{0} \beta^{\mu/2}},\label{eq:f}
\end{equation}
where $\beta$ is the ratio of gas pressure to the total pressure (The MN prescription hereafter). With expressions
(\ref{eq:visg}) and (\ref{eq:f}) for the viscous stresses and dissipated power
respectively, we can construct the basic equations describing the disc.
The azimuthal component of the equation of motion gives
\begin{equation}\label{eq:Phidirec}
8\pi\alpha_{0} H (p_{\rm gas})^{\mu/2} p^{(2-\mu)/2}= 3\Omega_{\rm K} \dot{M} J(R),
\end{equation}
where $J(R)=1-\sqrt{R_{\rm in}/R}$ and $R_{\rm in}$ denotes the
inner boundary of disc. Since we are interested in the regions of
the disc with radii much larger than $R_{\rm in}$, we have
$J\approx 1$, but we keep the more general expression in what follows.
Energy conservation implies that
\begin{equation}\label{eq:energy}
\sigma T_{\rm eff}^{4}=\frac{3}{8\pi}\Omega_{\rm K}^{2}\dot{M} J(R) (1-f).
\end{equation}
Finally, with the vertical transport of heat dominated by radiative diffusion
we have a relation between the midplane and surface temperatures given by
\begin{equation}
T = (\frac{3}{8}\kappa\Sigma)^{1/4} T_{\rm eff},\label{eq: T}
\end{equation}
where $\kappa$ is the opacity coefficient.

Equations (\ref{eq:zcom}), (\ref{eq:Phidirec}), (\ref{eq:energy}) and
(\ref{eq: T}) enable us to find $p$
and $T$ and $\rho$ as functions of $R$ with $\beta$ and $\zeta$ the critical input
parameters. Thus,
\begin{displaymath}
T=(\frac{4\sigma \Omega_{\rm K}}{3\kappa \alpha_{0}})^{-1/2}(\frac{16\pi^{2}\alpha_{0}^{2}ck_{\rm B}}
{3\sigma \mu_{\rm m} m_{\rm H} \dot{M}^{2}\Omega_{\rm K}^{4}J^{2}})^{-1/3}\frac{1}{(1+\zeta)^{1/6}}
\end{displaymath}
\begin{equation}\label{eq:main1}
\times \frac{(1-\sqrt{2\alpha_{0}\beta^{\mu/2}})^{1/2}}{
(1-\beta)^{1/3}}\beta^{(4-\mu)/12},
\end{equation}
\begin{displaymath}
p=(\frac{4\sigma \Omega_{\rm K}}{3\kappa \alpha_{0}})^{-1/2}(\frac{16\pi^{2}\alpha_{0}^{2}ck_{\rm B}}
{3\sigma \mu_{\rm m} m_{\rm H} \dot{M}^{2}\Omega_{\rm K}^{4}J^{2}})^{-2/3}(1+\zeta)^{1/6}
\end{displaymath}
\begin{equation}\label{eq:main2}
\times\frac{(1-\sqrt{2\alpha_{0}\beta^{\mu/2}})^{1/2}}{
(1-\beta)^{2/3}}\beta^{(8-5\mu)/12}.
\end{equation}
\begin{displaymath}
\rho=(\frac{8 \pi \alpha_{0}}{3 \Omega_{\rm k}^{2}\dot{M} J})^{2}  \frac{p^{3}}{(1+\zeta)} \beta^{\mu}.
\end{displaymath}

There are two algebraic equations for $\beta$ and $\zeta$ as follows
\begin{displaymath}
\frac{k_{\rm B}}{\mu_{\rm m} m_{\rm H}} (\frac{4\sigma\Omega_{\rm K}}{3\kappa \alpha_{0}})^{-3/2}
(\frac{8\pi \alpha_{0}}{3\Omega_{\rm K}^{2}\dot{M}  J})^{2} (\frac{16\pi^{2}\alpha_{0}^{2}c k_{\rm B}}
{3\sigma\mu_{\rm m} m_{\rm H} \dot{M}^{2}\Omega_{\rm K}^{4} J^{2}})^{-5/3}
\end{displaymath}
\begin{equation}\label{eq:beta}
 \times (1+\zeta)^{-5/6} \frac{
(1-\sqrt{2\alpha_{0}\beta^{\mu/2}})^{3/2}}{(1-\beta)^{5/3}}\beta^{(8+\mu)/12}=1,
\end{equation}
\begin{displaymath}
\zeta(1+\zeta)^{1/2}=\frac{4 \pi G}{\Omega_{\rm k}^{2}}(\frac{4\sigma\Omega_{\rm K}}
{3\kappa \alpha_{0}})^{-3/2}(\frac{8\pi\alpha_{o}}
{3 \Omega_{\rm k}^{2}\dot{M}j})^{2}.
\end{displaymath}
\begin{equation}\label{eq: zeta}
\times (\frac{16\pi^{2}\alpha_{0}^{2}c k_{\rm B}}{3\sigma\mu_{\rm m} m_{\rm H}\dot{M}^{2}\Omega_{\rm K}^{4}
 J^{2}})^{-2}\frac{(1-\sqrt{2\alpha_{0}\beta^{\mu/2}})^{3/2}}{(1-\beta)^{2}}\beta^{(8-\mu)/4},
\end{equation}
where $\mu_{\rm m}$ is the mean particle mass in units of the hydrogen atom mass, $m_{\rm H}$.
The other constants have their usual meaning. These two equations give us $\zeta$ as follows
\begin{displaymath}
\zeta=\frac{4 \pi G}{\Omega_{\rm k}^{2}}(\frac{4\sigma\Omega_{\rm K}}{3\kappa \alpha_{0}})^{-3/5}(\frac{8\pi\alpha_{o}}
{3 \Omega_{\rm k}^{2}\dot{M}j})^{4/5}(\frac{16\pi^{2}\alpha_{0}^{2}c k_{\rm B}}
{3\sigma\mu_{\rm m} m_{\rm H}\dot{M}^{2}\Omega_{\rm K}^{4}
 J^{2}})^{-1}
\end{displaymath}
\begin{equation}\label{eq: zzeta}
\times(\frac {k_{\rm B}}{\mu m_{\rm H} })^{-3/5}(1-\sqrt{2\alpha_{0}\beta^{\mu/2}})^{3/5}
\frac{\beta^{(16-3 \mu)/10}}{(1-\beta)}
\end{equation}
 Now equation (\ref{eq: zzeta}) enables us to find $T$, $P$ and $\rho$ as functions of $R$, $\beta$
and the other input parameters in the form
\begin{displaymath}
 T=(\frac{4\sigma \Omega_{\rm K}}{3\kappa \alpha_{0}})^{-1/5}(\frac{8\pi \alpha_{0}}
{3\Omega_{\rm K}^{2}\dot{M}  J})^{-2/5}(\frac {k_{\rm B}}{\mu m_{\rm H} })^{-1/5}
\end{displaymath}
\begin{equation}
\times (1-\sqrt{2\alpha_{0}\beta^{\mu/2}})^{1/5}\beta^{(2-\mu)/10},
\end{equation}
\begin{displaymath}
p=(\frac{4\sigma \Omega_{\rm K}}{3\kappa \alpha_{0}})^{-4/5}(\frac{8\pi \alpha_{0}}
{3\Omega_{\rm K}^{2}\dot{M}  J})^{2/5}(\frac {k_{\rm B}}{\mu m_{\rm H} })^{1/5}
\end{displaymath}
\begin{equation}\label{eq:main2}
\times(\frac{16\pi^{2}\alpha_{0}^{2}ck_{\rm B}}
{3\sigma \mu_{\rm m} m_{\rm H} \dot{M}^{2}\Omega_{\rm K}^{4}J^{2}})^{-1}
\frac{(1-\sqrt{2\alpha_{0}\beta^{\mu/2}})^{4/5}}{
(1-\beta)}\beta^{(4-2\mu)/5},
\end{equation}

\begin{displaymath}
\rho=(\frac{4\sigma \Omega_{\rm K}}{3\kappa \alpha_{0}})^{-3/5}(\frac{8\pi \alpha_{0}}
{3\Omega_{\rm K}^{2}\dot{M}  J})^{4/5}(\frac {k_{\rm B}}{\mu m_{\rm H} })^{-3/5}
\end{displaymath}
\begin{equation}\label{eq:main2}
\times(\frac{16\pi^{2}\alpha_{0}^{2}ck_{\rm B}}
{3\sigma \mu_{\rm m} m_{\rm H} \dot{M}^{2}\Omega_{\rm K}^{4}J^{2}})^{-1}
\frac{(1-\sqrt{2\alpha_{0}\beta^{\mu/2}})^{3/5}}{
(1-\beta)}\beta^{(16-3\mu)/10}.
\end{equation}

In order to study the behavior of our solutions, it is more convenient
to introduce dimensionless variables. For the central mass $M$,  we
introduce $M_8=M/(10^{8}M_{\odot})$ and for the radial distance $R$,
we have $r_3=R/(10^{3}R_{\rm S})$, where $R_{\rm S}=2GM/c^{2}$ is
the Schwarzschild radius. The mass accretion rate can be written as
\begin{equation}
\dot{M}=\frac{l_{\rm E}}{\epsilon}\frac{4\pi GM}{\kappa_{\rm e.s.}c}=\frac{l_{\rm E}}{\epsilon}\frac{L_{\rm E}}{c^{2}},
\end{equation}
where $l_{\rm E}=L/L_{\rm E}$ is the dimensionless disc luminosity relative to the Eddington limit,
$\epsilon=L/(\dot{M}c^{2})$ is the radiative efficiency and $\kappa_{\rm e.s.}\approx 0.04$
$\rm m^{2} kg^{-1}$ is the electron opacity. In our analysis, we will use the nondimensional factor
$l_{\rm E}/\epsilon$ as a free parameter so that by changing this parameter we can consider
appropriate values of the accretion rate. However, some authors introduce different forms for
the accretion rate. For example, Nayakshin \& Cuadra (2005) who studied gravitational stability
of the Galactic Center, introduced $\dot{M}= (\dot{m}/\epsilon)(L_{\rm E}/c^{2})$  with
$\epsilon \approx 0.06$ and $\dot{m}=0.03$ to $1$. These values, which are appropriate for the
Galactic Center, correspond to $l_{\rm E}/\epsilon \approx 0.5$ to $16.6$ in our notation. Also,
for a central mass with mass $M=10^{8} M_{\odot}$, Goodman \& Tan (2004) proposed $l_{\rm E}/\epsilon = 10$.
Thus, in our analysis, the corresponding range of values for $l_{\rm E}/\epsilon$ is from $1$ to $10$.

Our solutions for $T$, $P$ and $\rho$ become
\begin{displaymath}
T=1.65\times10^{4} \alpha_{0}^{-1/5}\hat{\kappa}^{1/5} M_{8}^{-1/5}(\frac{l_{\rm E}}{\epsilon})^{2/5} J^{2/5} r_{3}^{-9/10}
\end{displaymath}
\begin{equation}\label{eq:Tem}
 \times \beta^{(2-\mu)/10}(1-\sqrt{2\alpha_{0}\beta^{\mu/2}})^{1/5},\label{eq:p}
\end{equation}

\begin{displaymath}
p=18.72\times \alpha_{0}^{-4/5}\hat{\kappa}^{4/5} M_{8}^{-4/5}(\frac{l_{\rm E}}{\epsilon})^{8/5} J^{8/5} r_{3}^{-18/5}
\end{displaymath}
\begin{equation}
 \times \beta^{(4-2\mu)/5}(1-\beta)^{-1}(1-\sqrt{2\alpha_{0}\beta^{\mu/2}})^{4/5},\label{eq:p}
\end{equation}
\begin{displaymath}
 \rho=0.823\times10^{-7} \alpha_{0}^{-3/5}\hat{\kappa}^{3/5} M_{8}^{-3/5}
(\frac{l_{\rm E}}{\epsilon})^{6/5} J^{6/5} r_{3}^{-27/10}
\end{displaymath}
\begin{equation}
\times \beta^{(16-3\mu)/10}(1-\beta)^{-1}(1-\sqrt{2\alpha_{0}\beta^{\mu/2}})^{3/5},\label{eq:rho}
\end{equation}

\begin{displaymath}
 \frac{H}{R}=6.7\times10^{-3}\alpha_{0}^{-1/5}\hat{\kappa}^{-4/5} M_{8}^{-1/5}
(\frac{l_{\rm E}}{\epsilon})^{-3/5} J^{-3/5}r_{3}^{11/10}
\end{displaymath}
\begin{equation}
 \times \beta^{-(8+\mu)/10}(1-\beta)(1-\sqrt{2\alpha_{0}\beta^{\mu/2}})^{-4/5},\label{eq:HR}
\end{equation}
We can also calculate the surface density as
\begin{displaymath}
\Sigma = 3.292\times 10^{5} \alpha_{0}^{-4/5}\hat{\kappa}^{-1/5} M_{8}^{1/5}
(\frac{l_{\rm E}}{\epsilon})^{3/5} J^{3/5} r_{3}^{-3/5}
\end{displaymath}
\begin{equation}
\times \beta^{(4-2\mu)/5}(1-\sqrt{2\alpha_{0}\beta^{\mu/2}})^{-1/5}.
\end{equation}
and the ratio $\beta$ is obtained from nondimensional form of equations (\ref{eq:beta}) and (\ref{eq: zzeta}), i.e.
\begin{displaymath}
 \zeta =0.134\times \alpha_{0}^{-3/5}\hat{\kappa}^{3/5} M_{8}^{7/5}(\frac{l_{\rm E}}{\epsilon})^{6/5} J^{6/5} r_{3}^{3/10}
\end{displaymath}
\begin{equation}\label{eq:zeta2}
\times \beta^{(16-3\mu)/10}(1-\beta)^{-1}(1-\sqrt{2\alpha_{0}\beta^{\mu/2}})^{3/5}\,\label{eq:beta1}
\end{equation}
\begin{displaymath}
0.216\times \alpha_{0}^{1/5}\hat{\kappa}^{9/5} M_{8}^{1/5}(\frac{l_{\rm E}}{\epsilon})^{8/5} J^{8/5} r_{3}^{-21/10}
\end{displaymath}
\begin{equation}\label{eq:beta2}
\times \beta^{(8+\mu)/10}(1-\beta)^{-2}(1-\sqrt{2\alpha_{0}\beta^{\mu/2}})^{9/5}-1-\zeta=0\,\label{eq:beta1}
\end{equation}
where $\hat{\kappa}=\kappa/\kappa_{\rm e.s.}$ and we assumed $\mu_{\rm m} = 0.6$.

Equations (\ref{eq:Tem}), (\ref{eq:p}), (\ref{eq:rho}) and (\ref{eq:HR})  describe the structure of
a self-gravitating disc with a
dissipative corona. However, the physical variables depend not only
on the radial distance but also on $\beta$ and $\zeta$ which can be calculated at each radius from algebraic
equations (\ref{eq:zeta2}) and (\ref{eq:beta2}).  In the next section we
will analyze our solution, in particular the gravitational stability
of the disc.

\section{Analysis}

We can  solve  algebraic equations  (\ref{eq:zeta2}) and
(\ref{eq:beta2}) numerically at each radius to obtain  the ratios
$\beta$ and $\zeta$. Having these parameters, physical variables are
obtained   based on our analytical solutions. Since we are
interested in gravitational stability of the disc, we can approximate
$J(R)\simeq 1$ for $R\gg R_{\rm in}$. Also we fix the central mass
$M_{8}=1$, the opacity $\hat{\kappa}=1$ and the viscosity exponent
$\mu=1$, but vary the other input parameters.  However, for the
transport of energy from the disc to the corona (i.e. the fraction
$f$ in our approach),  we may either consider the MN prescription (in
which $f$ is not constant) or assume that the fraction $f$ is
constant and use it as a free parameter of the model. We study
properties of the solutions in both cases in the next two
subsections. Interestingly,  typical behaviors of the solutions are
similar qualitatively, though precise values of the physical
variables at each radius depend on the mathematical prescription of
the fraction $f$. The possible effects of the self-gravity of the
disc in both forms of prescription for the fraction $f$ are
illustrated by appropriate plots in the next subsections.

\subsection{Properties of the solutions corresponding to $f=\sqrt{2\alpha_{0} \beta^{\mu/2}}$}

\begin{figure*}
\vspace*{-150pt}
\epsfig{figure=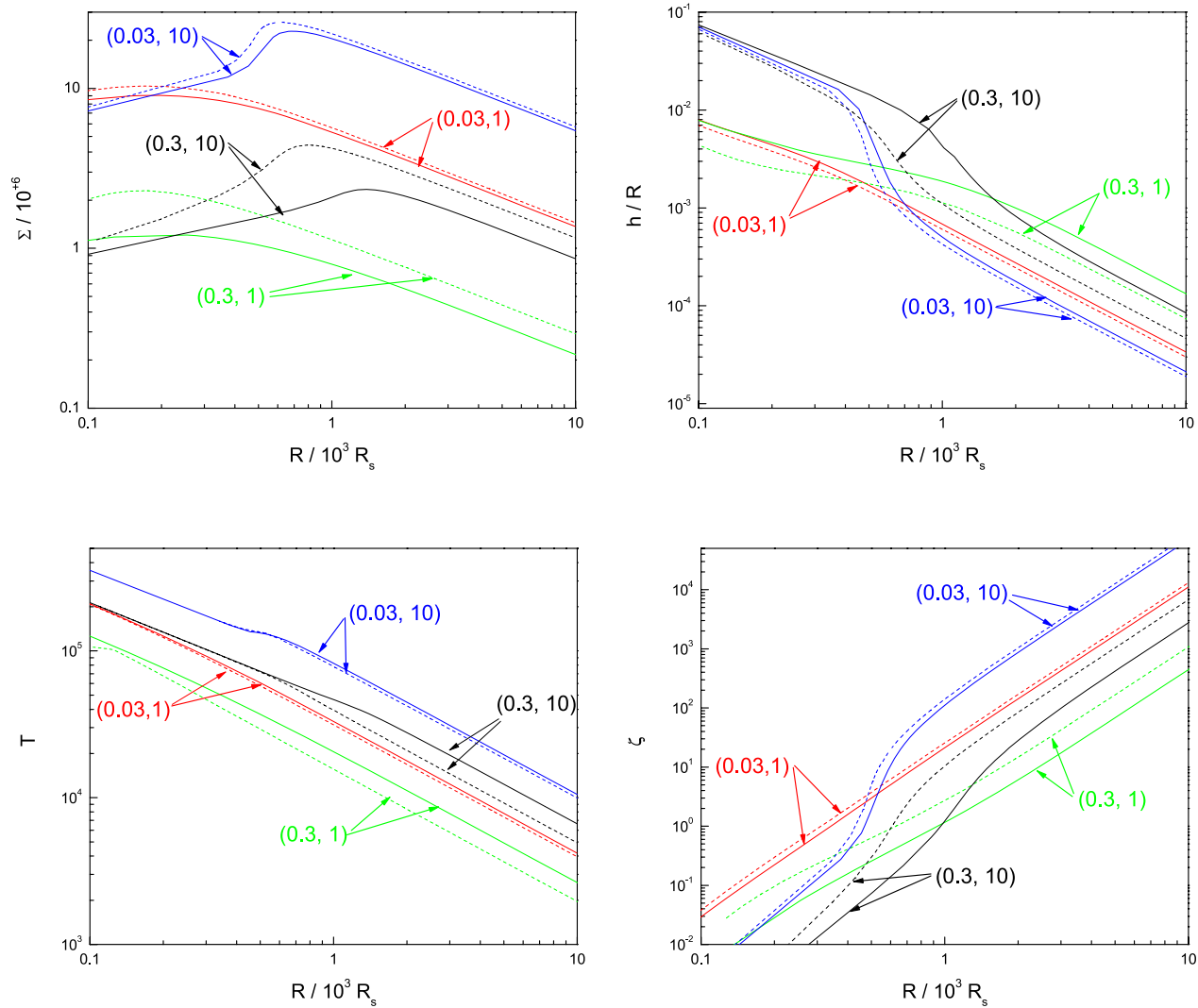,angle=0,scale=0.9} \caption{Profiles of the
physical variables  for a disc without a corona ({\it solid lines})
and with a corona ({\it dashed lines})  vs. radial location in the
disc  with $\mu=1.0$, $\mu_{\rm m}=0.6$, $\hat{k}=1$ and $M_{8}=1$.
Each curve is  labeled  by the viscosity coefficient $\alpha_0$  and
the accretion rate $l_{\rm E}/\epsilon$ as a pair $(\alpha_{0},
l_{\rm E}/\epsilon)$. A different color is used
for the curve corresponding  to each pair of the input parameters. Here, the MN
prescription for the dissipated energy into the corona has been
considered. Since the amount of  energy transported into the
corona increases with the viscosity coefficient, the possible
effects of the corona on the profiles are more evident for the
higher viscosity coefficient. While the temperature of the disc
decreases  because of the corona, the surface density of the disc
increases.}\label{fig:figure1}
\end{figure*}

\subsubsection{General properties}

Figure \ref{fig:figure1} shows the typical behaviours of the solutions  for discs with (dashed curves) and
without (solid curves) a corona.  Each curve is labeled  by the appropriate values of
$(\alpha_{0},l_{E}/\epsilon)$. Behavior of the surface density
is shown in the  top left-hand plot of Figure \ref{fig:figure1}. For the same input parameters,
the presence of a corona enhances the surface density, and the surface density generally
displays two regimes in the inner and outer regions. The size of the
inner part is sensitive to the values of the input parameters.
For a fixed viscosity coefficient, when the accretion rate increases  the surface density in
the inner part decreases, while the surface density at the outer parts increases in both cases with
and without a corona. For a fixed accretion rate when the viscosity coefficient increases, the
surface density decreases. Generally, we see that the corona becomes more effective when the
viscosity and the accretion rate have larger values. Finally,  at large radii the surface  density
is in proportion to  $ R^{-0.6}$.

The top right-hand plot of Figure \ref{fig:figure1} shows the ratio of disc thickness to the
radius. The corona causes the disc to become thinner in comparison with the same disc without a corona.
In the absence of a corona, for a fixed rate of accretion, viscosity increases the disc thickness.
However, in a disc with a corona, so long as the accretion rate is constant, the
thickness of the disc decreases in the inner part and increases in the outer part due to increase of the viscosity
coefficient. With a fixed  viscosity coefficient and  increased accretion rate,
the thickness of the disc increases in the inner part, but  the thickness of the outer part decreases
in both the case of a disc with a corona and the case without a corona. The disc thickness scales like
$ R^{-1.3}$ at large radii.

 The bottom left-hand plot of Figure \ref{fig:figure1} shows that the presence of a corona serves to
cool the disc because of the energy transported from the disc to the corona.
Generally, for a fixed accretion rate when
the  viscosity coefficient increases, the disc temperature decreases because more energy is transported from the disc to the corona. But  for a fixed  viscosity coefficient when the accretion rate increases, then the  temperature of the disc increases too.
At large radii the  temperature is changing in proportion to $R^{-0.9}$.

The bottom right-hand plot of Figure \ref{fig:figure1} shows that the corona causes the disc
to become more self-gravitating in the vertical direction in comparison with the case without a corona
while, for a fixed accretion rate, when the viscosity coefficient increases, the self-gravity of the disc in
the vertical direction decreases. However, for a fixed viscosity coefficient, when the accretion rate
increases,  there are two regimes, in  both the case with and without a corona. At smaller radii,
self-gravity of the disc in the vertical direction decreases, but as we go to the larger radii
the vertical component of self-gravity becomes more significant. At larger radii the ratio $\zeta$
varies in proportion to $R^{2.75}$. This plot shows that the vertical component of the self-gravity
can be comparable to the vertical component of the gravitational force of the central object at radii that are not large. Clearly, this radius is within a distance from the central object where the corona can
exist. For instance for a pair $(0.03, 1)$,  the self-gravity of the disc is of the order of the gravitational
force of the central mass at around $R \sim 300 R_{s}$. For the $(0.03, 10)$ case this  happens at  around
$R \sim 400 R_{s}$. It clearly shows that self-gravity in the vertical direction is important for
a disc-corona system even at not large radii from the central object where a corona exists.

\begin{figure*}
\vspace*{-150pt}
\epsfig{figure=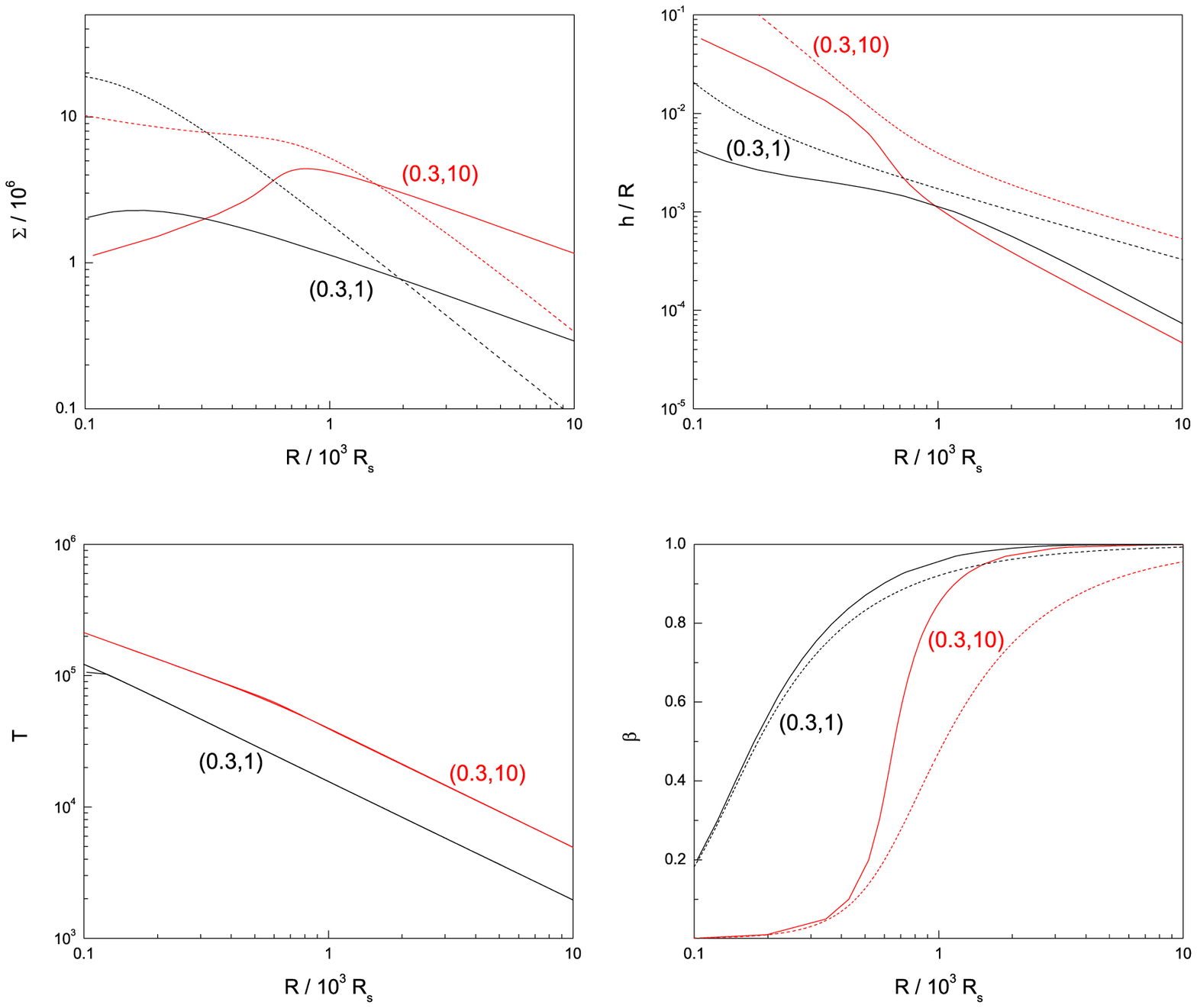,angle=0,scale=0.9}
\caption{Effects of the self-gravity of the disc on the profiles of
the physical variables are shown in these plots by comparing our
solutions with a corona ({\it solid lines}) and those solutions with
a corona but neglecting the vertical component of the self-gravity
of the disc (KS solutions, {\it dashed lines}). The input parameters
are the same as Figure \ref{fig:figure1}. Each curve is labeled  by
a pair $(\alpha_{0}, l_{\rm E}/\epsilon)$. Different colors  are used to denote the curves corresponding to the different pair of  input parameters.
Here, the MN prescription for the dissipated energy into the corona
has been considered. We see reduction in the surface density of the disc at radii approximately
smaller than $10^{3} R_{s}$ due to the self-gravity of the disc. Beyond this radius the surface
density increases because of the self-gravity of the disc. However, the
temperature of the disc is independent of the self-gravity of the
disc.}\label{fig:figure2}
\end{figure*}
\begin{figure*}
\vspace*{-150pt}
\epsfig{figure=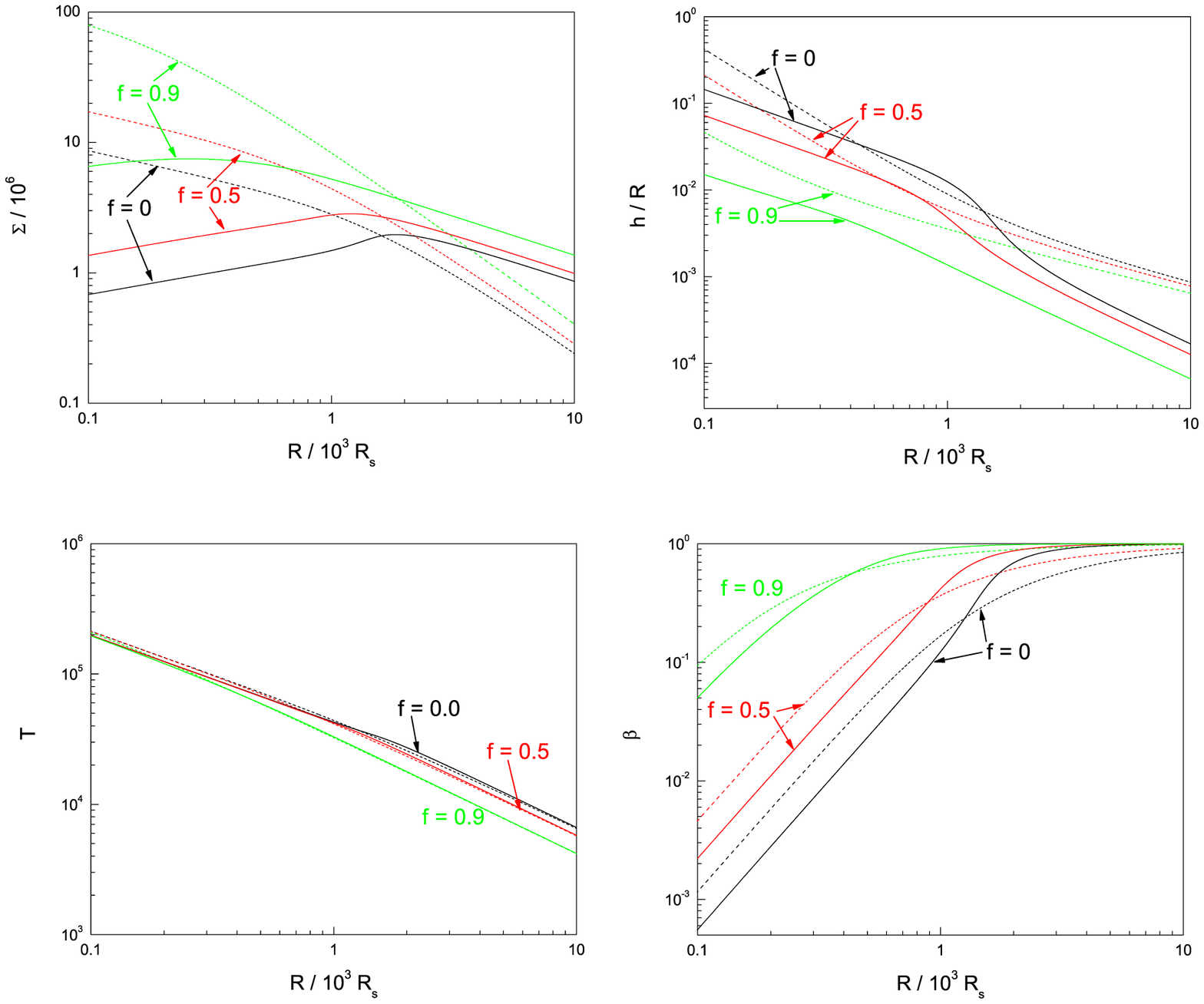,angle=0,scale=0.9} \caption{Physical profiles
of  the disc when the fraction of dissipated energy into the corona
is constant. The input parameters are $\mu=1.0$, $\mu_{\rm m}=0.6$,
$\hat{k}=1$, $\alpha_{0}=0.3$, $l_{\rm E}/\epsilon =10$ and
$M_{8}=1$. Each curve is labeled by the fraction $f$ (which is
assumed to be constant) and is shown by different color. {\it Solid
lines} are for the solutions with the vertical component of the
self-gravity of the disc and the {\it dashed lines} show solutions
when self-gravity of the disc is neglected. These curves are showing
similar behaviors to when the fraction $f$ is described by the MN
prescription (Figures \ref{fig:figure1} and \ref{fig:figure2}).
Existence of the corona lead to an increase of the surface density
and a cooler disc. Also, the surface density in the inner parts of the
disc reduces because of the vertical component of the self-gravity of
the disc, though the surface density increases beyond these
regions.}\label{fig:figure3}
\end{figure*}

\subsubsection{The effect of the self-gravity of the disc}

We study the possible effects of the self-gravity of the disc on the profiles of the physical variables of the disc in this subsection. Gravitational stability of the disc and the rule of the self-gravity is studied in Section 4 separately. Figure \ref{fig:figure2} compares profiles of the solutions with self-gravity with those without self-gravity (KS solutions). The input parameters are similar to Figure \ref{fig:figure1} and each curve is labeled by a pair $(\alpha_{0}, l_{\rm E}/\epsilon)$. We see that the surface density at radii smaller than $10^3$  Schwarzschild radius decreases due to the effect of self-gravity of the disc, irrespective of the viscosity coefficient and the accretion rate. Although the fraction of $\zeta$ is not very large at these radii, the profile of the surface density changes significantly because of the self-gravity of the disc.

Also, the self-gravity of the disc implies a thinner disc with a smaller opening angle. This reduction in the thickness of the disc is at all radii and for  all the input parameters. In our analysis, we do not consider the vertical structure of the accretion disc. But in  studies of the vertical structure of the accretion discs and analysis of the energy transport in this direction (e.g., via convection), one should note that the self-gravity of the disc significantly reduces the thickness of the disc according to our analytical solutions for a disc and corona system. However, the temperature of the disc is independent of the self-gravity of the disc for all the input parameters. However,  the ratio of the gas pressure to the total pressure, in particular for high accretion rates, increases because of the self-gravity of the disc.

\begin{figure}
\vspace*{+37pt}
\epsfig{figure=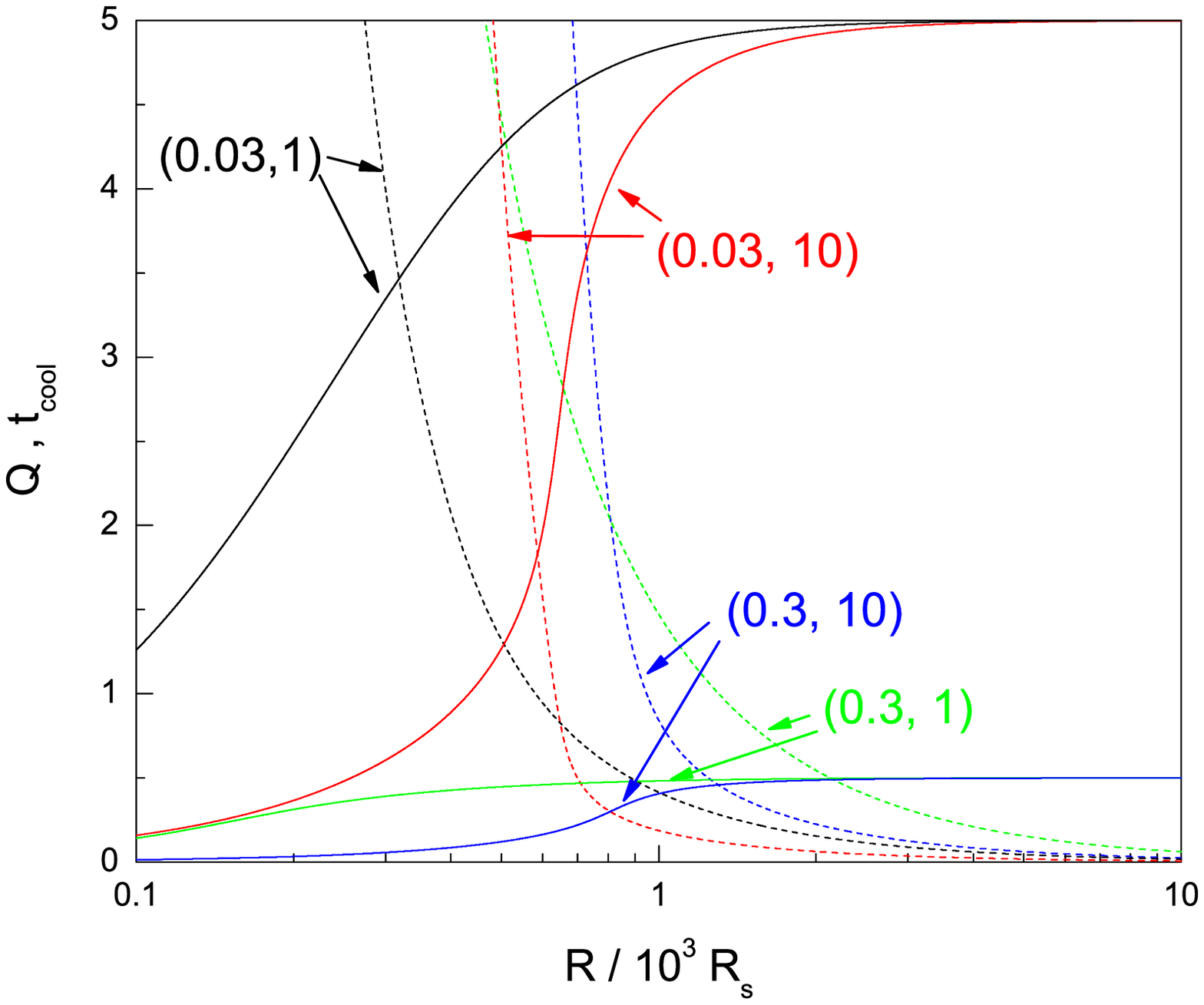,angle=0,width=10cm} \caption{Profiles of the
cooling time-scale $t_{\rm cool}$ ({\it solid lines}) and the Toomre
parameter $Q$ ({\it dashed lines}) vs. radial location in the disc.
The curves are corresponding to the MN prescription and the input
parameters are $\mu=1.0$, $\mu_{\rm m}=0.6$, $\hat{k}=1$, $\gamma=2$
and $M_{8}=1$.  Each curve is labeled by a pair
$(\alpha_{0}, l_{\rm E}/\epsilon)$ and is showing by a
color.}\label{fig:figure4}
\end{figure}

\begin{figure}
\vspace*{+20pt}
\epsfig{figure=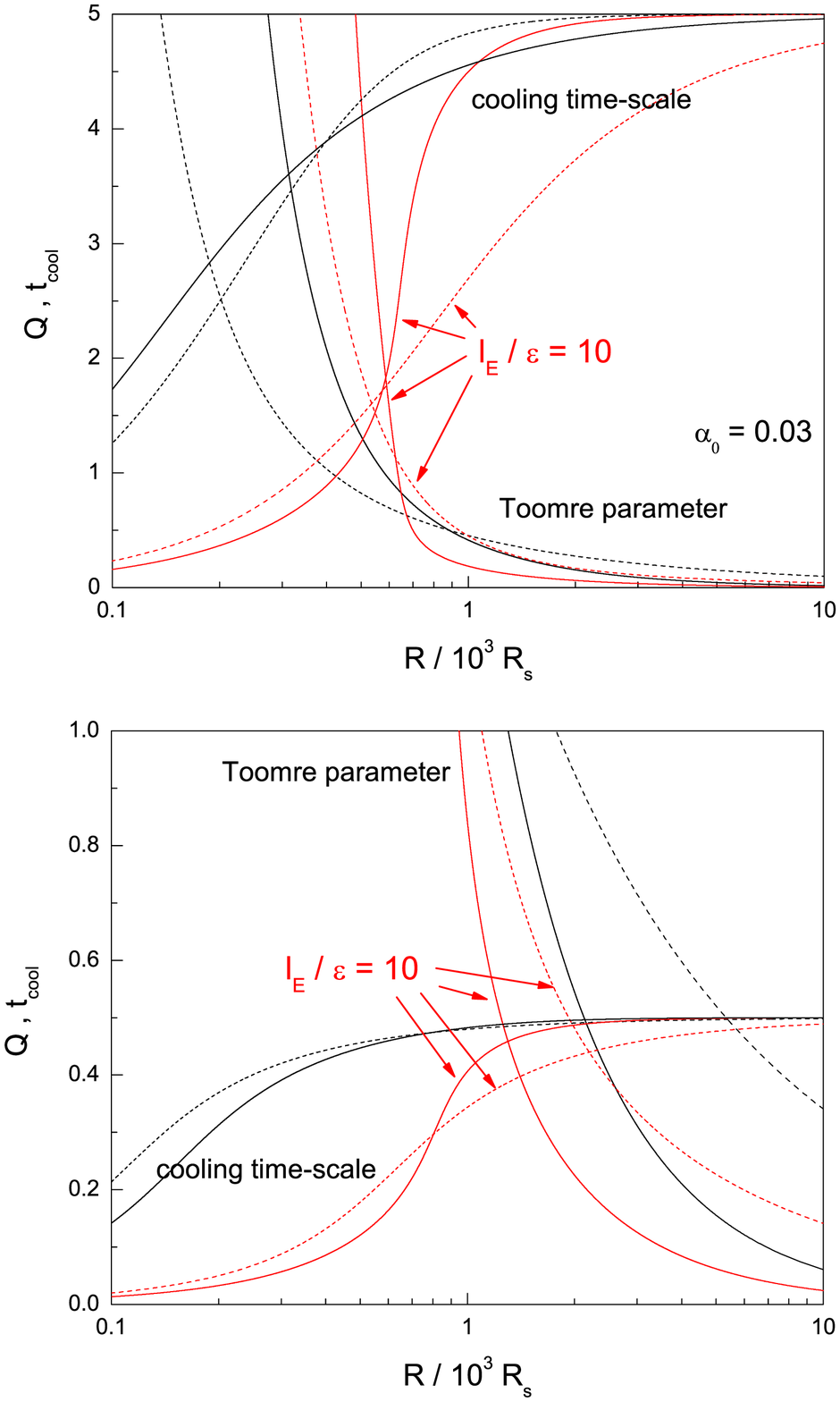,angle=0,width=10.0cm} \caption{Profiles of the
cooling time-scale $t_{\rm cool}$ and the Toomre parameter $Q$  vs.
radial location in the disc for $\alpha_{0}=0.03$ ({\it top}) and
$\alpha_{0}=0.3$ ({\it bottom}). Solid curves are corresponding to
our solutions, but dashed lines are for solutions neglecting
self-gravity of the disc (KS solutions). The curves are
corresponding to the MN prescription and the input parameters are
$\mu=1.0$, $\mu_{\rm m}=0.6$, $\hat{k}=1$, $\gamma=2$ and $M_{8}=1$.
Curves corresponding to $ l_{\rm E}/\epsilon = 10$ are marked and
the rest of the curves are for  $ l_{\rm E}/\epsilon =
1$.}\label{fig:figure5}
\end{figure}

\subsection{Properties of the solutions when $f$ is a free parameter}

Svensson \& Zdziarski  (1994) presented analytical solutions for the disc-corona systems, in which a constant fraction $f$ is dissipated into the corona. Now, we analysis properties of the physical variables of those solutions that correspond to a constant fraction $f$.  Figure \ref{fig:figure3} shows profiles for $\mu=1.0$, $\mu_{\rm m}=0.6$, $\hat{k}=1$, $\alpha_{0}=0.3$, $l_{\rm E}/\epsilon =1$ and $M_{8}=1$. Each curve is labeled by the fraction $f$, which is assumed to be constant. For easier comparison, solutions without self-gravity are shown by  dashed lines. Interestingly, properties of the solutions are qualitatively similar to the solutions with the MN prescription.  Here, the surface density of the disc decreases due to the self-gravity of the disc at radii smaller than $10^3$ Schwarzschild radius. This reduction is independent of the amount of the energy transported to the corona. However, as the fraction $f$ increases, the profile of the surface density shifts upward either with or without self-gravity of the disc. This means that as more energy is transported via magnetic tubes from the disc to the corona, the surface density increases.

The thickness of the disc  also depends on the self-gravity of the disc and the fraction $f$. The disc becomes thinner when more energy is transferred from the disc to the corona. Also, for a given fraction $f$, self-gravity of the disc causes the thickness of the disc to decrease. The thickness of the disc corresponding to the solutions with self-gravity is more sensitive to the fraction $f$  compared with the solutions without self-gravity. Again, in this case, the existence of the corona implies a cooler disc, but the temperature of the disc is independent of the self-gravity of the disc. The ratio $\beta$ also decreases at the inner parts of the disc due to the self-gravity of the disc. However, this reduction is more evident for a smaller fraction $f$.

\section{Gravitational stability of the disc-corona system}

Having our analytical solutions for the steady-state structure of an accretion disc with a corona, we can study the gravitational stability of such a system. Although most authors have been using the Toomre parameter as the main criterion in order to address the gravitational stability of accretion discs (e.g., Goodman 2003; Goodman \& Tan  2004; Nayakshin \& Cuadra 2005), recent numerical simulations and physical considerations  show that the cooling time-scale is  another important physical factor (e.g., Gammie 2001; Lodato \& Rice 2004). Also, the analysis of KS of the gravitational stability of an accretion disc with corona is based on just the Toomre parameter (they  had also neglected self-gravity of the disc). In this subsection, we study gravitational stability of our solutions considering the Toomre parameter and the cooling time-scale. Since solutions with free $f$ and the MN prescription show qualitatively  similar behaviors, the stability analysis is restricted to those corresponding to the solutions with the MN prescription. First, we calculate these parameters, then we do a parameter study.

\subsection{Toomre Parameter}

It was shown by Toomre (1964) that a rotating disc is
subject to gravitational instabilities when the Q parameter
\begin{equation}
Q =\frac{c_{s}\Omega}{\pi G \Sigma},
\end{equation}
becomes smaller than a critical value, which is close to unity,
where $c_{s}$ is the sound speed inside the accretion disc and
$\Omega =\Omega_{K}$ is the angular velocity. Having our analytical
solution for the system,  Toomre parameter of the model becomes
\begin{displaymath}
Q=4.945\times \alpha_{0}^{7/10}\hat{\kappa}^{3/10}
M_{8}^{-13/10}(\frac{l_{\rm E}}{\epsilon})^{-2/5} J^{-2/5}
r_{3}^{-27/20}
\end{displaymath}
\begin{equation}
\times \beta^{(7
\mu-24)/20}(1-\sqrt{2\alpha_{0}\beta^{\mu/2}})^{3/10}.\label{eq:QQ}
\end{equation}
This equation along with two algebraic equations (\ref{eq:zeta2})
and (\ref{eq:beta2}) give $Q$ as a function of the radial distance.
Generally, $Q$ is much greater than unity in the inner parts of the
disc which implies these regions are gravitationally stable and do
not fragment. But the Toomre parameter decreases with increasing
radial distance so that $Q$ reaches the critical value of unity at a
self-gravitating radius which we denote  it by $R_{\rm  sg}$. Thus, all
regions with $R>R_{\rm sg}$ are gravitationally unstable and may
fragment. However, this is only true as long as the cooling time-scale  also satisfies a criterion which will be explored in the next subsection.

It can be shown that when the Toomre parameter is of order unity the
ratio $\zeta$ is around $4.8$ in our model. With $Q =1$, we have
\begin{equation}
 4 \pi G \rho=\frac{2 c_{s}\Omega_{k}}{H}.
\end{equation}
When combined with equation (\ref{eq:zetaf}) this gives
\begin{equation}
\zeta=\frac{2 c_{s}}{H \Omega_{k}}.\label{eq:1}
\end{equation}
From (\ref{eq:zcom}) and $c_{s}=\sqrt{p/\rho}$, we have
\begin{equation}
\frac{c_{s}}{ H \Omega_{k}}=\sqrt{1+\zeta}.\label{eq:2}
\end{equation}
Using equations (\ref{eq:1}) and (\ref{eq:2}), we obtain an
algebraic quadratic equation for $\zeta$. The only physical root of
this equation is $\zeta = 2(1+\sqrt{2})\simeq 4.8$. This means that  at a
radius where the Toomre parameters is around unity, we have
$\zeta\simeq 4.8$, i.e., the force of the self-gravity of the disc in
the vertical direction is $4.8$ times the  vertical component of the gravitational force of
the central object.

 The most unstable wavelength for the $Q \sim 1$ disc is of the order of the disc vertical scale height $H$ (Toomre
1964). Thus, the most unstable mode has radial wave number $k_{\rm
mu}=(QH)^{-1}$ and so the mass of a fragment at $R=R_{\rm sg}$
becomes
\begin{equation}
M_{\rm frag} \approx \Sigma (\frac{2\pi}{k_{\rm
mu}})^{2}=4\pi^{2}\Sigma H^{2},\label{eq:frag}
\end{equation}
which gives
\begin{displaymath}
 M_{\rm frag} =0.126 \times10^{34} \alpha_{0}^{1/5}\hat{\kappa}^{-6/5} M_{8}^{-4/5}
(\frac{l_{\rm E}}{\epsilon})^{-7/5} J^{-7/5} r_{3}^{-6/5}
\end{displaymath}
\begin{equation}
\times
\beta^{(\mu-32)/10}(1-\beta)^{2}(1-\sqrt{2\alpha_{0}\beta^{\mu/2}})^{-6/5}
. \label{eq:beta1}
\end{equation}

\begin{figure*}
\vspace*{-150pt}
\epsfig{figure=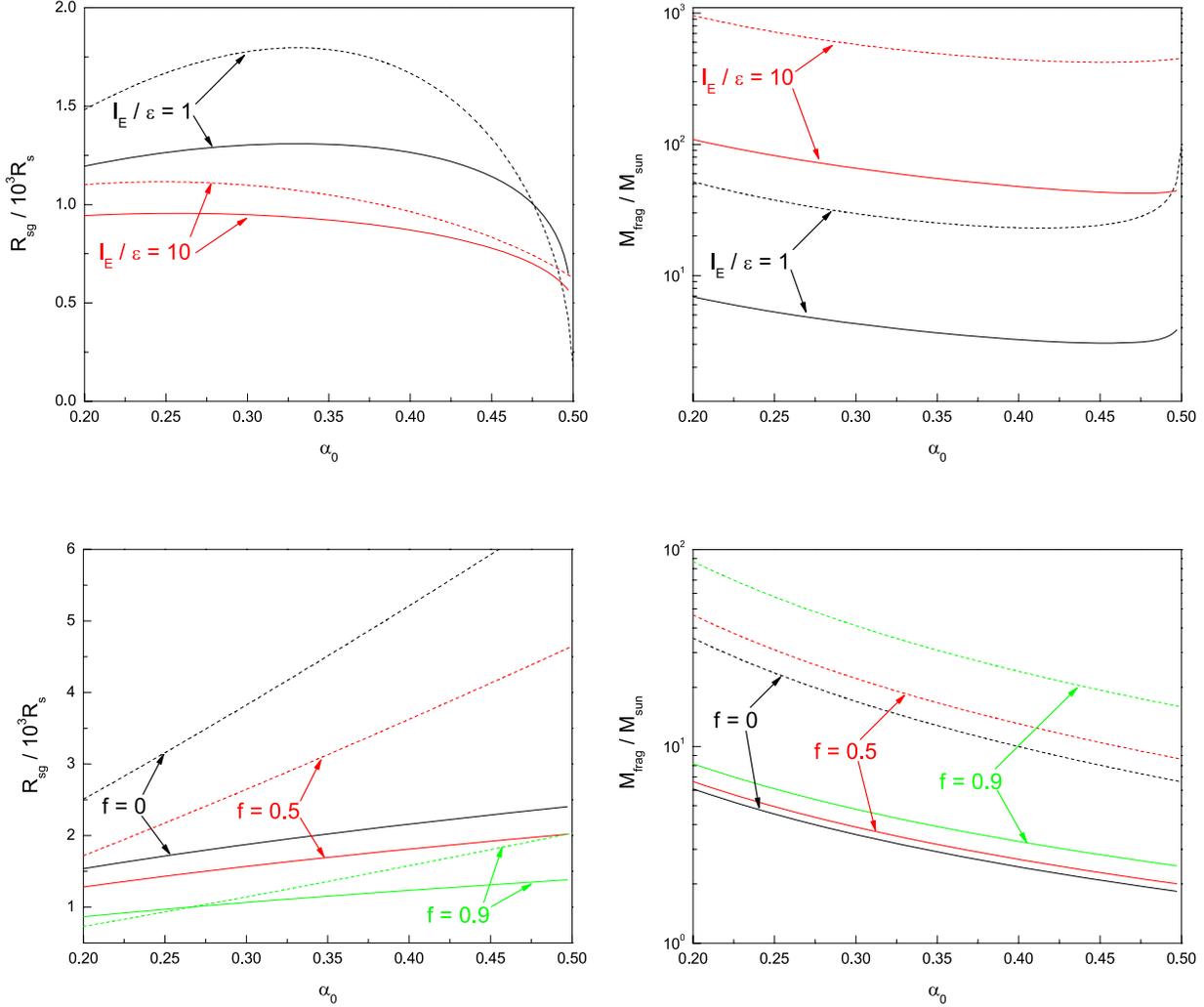,angle=0,scale=0.9} \caption{The self-gravitating radius $R_{\rm sg}$ (in Schwarzschild radius $R_{\rm S}$) and the mass
of the first clumps $M_{\rm frag}$ (in  solar mass) for $\mu=1.0$, $\mu_{\rm m}=0.6$, $\hat{k}=1$, and $M_{8}=1$. {\it Top} plots are corresponding to the solutions with MN prescription and {\it bottom} plots are for the solutions with fraction $f$ as a free parameter. Solutions with self-gravity are showing by solid curves, but dashed lines are corresponding to the solutions without self-gravity. For solutions with free $f$, each curve is labeled by this parameter and we assume $l_{\rm E}/\epsilon =1$.}\label{fig:figure6}
\end{figure*}

\subsection{Cooling time-scale}
There are actually different approaches to calculate the cooling
time-scale. We can estimate the cooling time-scale $\tau_{\rm cool}$
using simple cooling laws as the ratio of the internal energy density
$\epsilon=\Sigma c_{\rm s}^2 /\gamma (\gamma -1)$ to the cooling rate $\Lambda$, i.e. $\tau_{\rm
cool}=\epsilon/\Lambda$. Using equation (\ref{eq:energy}), we can write
\begin{equation}
\Lambda = (\frac{1}{1-f}) \sigma T_{\rm eff}^{4}=\frac{3}{8\pi}\Omega_{\rm K}^{2}\dot{M} J(R).\label{eq:cool1}
\end{equation}
On the other hand, angular momentum equation (\ref{eq:Phidirec}) gives
\begin{equation}
\alpha_{0}H (p_{\rm gas})^{\mu/2} p^{(2-\mu)/2}=\frac{3}{8\pi}\Omega_{\rm K}\dot{M}J(R).\label{eq:cool2}
\end{equation}
Having equation (\ref{eq:cool2}), we can rewrite equation (\ref{eq:cool1}) as
\begin{equation}
\Lambda = \alpha_{0} H \Omega_{\rm K} (p_{\rm gas})^{\mu/2} p^{(2-\mu)/2}.\label{eq:cool3}
\end{equation}
Thus, our cooling time-scale becomes
\begin{equation}
\tau_{\rm cool}=\frac{1}{\alpha_{0} \gamma (\gamma -1)}\frac{1}{\Omega_{\rm K}}(\frac{p_{\rm gas}}{p})^{(2-\mu)/2},
\end{equation}
and non-dimensional cooling time-scale becomes
\begin{equation}
t_{\rm cool}=\frac{\tau_{\rm cool}}{t_0}=\frac{1}{3\alpha_0 \gamma
(\gamma -1) } \beta^{(2-\mu)/2},\label{eq:cool}
\end{equation}
where $\gamma$ is the ratio of the specific heats and
$t_0=3/\Omega_{\rm K}$. According to Gammie (2001) and Lodato \&
Rice (2004) the fragmentation occurs when $t_{\rm cool} < 1$.

\subsection{Gravitational instability: A parameter study}
When the disc cools very rapidly, the self-gravity of the disc does not have
enough time to prevent the formation of bound objects in the disc.
Since a corona extracts energy from the disc, we think, existence of
the corona decreases the cooling time scale and the disc becomes
more prone to  fragmentation. As we have already mentioned, in the KS analysis of the
gravitational instability of the disc and corona system, only the Toomre
criteria has been used. Not only we consider this criterion of
 fragmentation, but  the cooling time-scale condition is also
considered in this paper.

Figure \ref{fig:figure4} shows the cooling time scale and the Toomre
parameter versus radial location in the disc. For the energy
exchange from the disc to the corona, the MN prescription is used and
the other input parameters are $\mu=1.0$, $\mu_{\rm m}=0.6$,
$\hat{k}=1$, $\gamma=2$ and $M_{8}=1$. Approximately, when $\gamma=5/3$ the cooling time-scale increases  by a factor of two in comparison to the case with $\gamma=2$.  Each curve is labeled by a pair $(\alpha_{0}, l_{\rm E}/\epsilon)$. Solid lines are for
$t_{\rm cool}$ and dashed lines correspond to the parameter
$Q$. In the inner parts of the disc, the Toomre parameter is very
large, irrespective of the input parameters. Although the cooling
time scale is less than unity (except for $(0.03, 1)$) in these
regions, the Toomre condition for fragmentation does not satisfy and so the
disc does not fragment. However, as we move toward larger radii,
the Toomre parameter decreases  down to the values smaller than unity. But the nondimensional cooling time-scale increases from small values at the inner parts of the disc to values greater or smaller than unity depending on the input parameters. Since in the MN prescription the amount of the transported energy from the disc to the corona is directly proportional to the viscosity coefficient $\alpha_0$, as this parameter increases, more energy is
transported from the disc into the corona and the cooling time-scale decreases. According to Figure \ref{fig:figure4} when the viscosity coefficient is $0.03$, the nondimensional cooling time-scale increases to values larger than unity at regions of the disc where the Toomre parameter is less than unity, irrespective of the accretion rate. When accretion rate decreases, the increase of $t_{\rm cool}$ becomes more smooth. However, for a larger viscosity coefficient $0.3$ the cooling time-scale is lower than unity even at larger radii. So, in this case, the disc can fragment at regions where the Toomre parameter reaches to values lower than unity.

Figure \ref{fig:figure5} shows possible effects of the self-gravity of the disc on the Toomre parameter and the cooling time-scale. We compare curves of $Q$ and $t_{\rm cool}$ for $\alpha_{0}=0.03$ (top plot) and $\alpha_{0}=0.3$ (bottom plot). While solid curves  correspond to our solutions, the dashed lines are for solutions neglecting the self-gravity of the disc (KS solutions). As we discussed, our solutions for $\alpha_0 = 0.03$ are gravitationally stable considering cooling time-scale and Toomre conditions. However, we see that while the self-gravity of the disc  slightly causes  $t_{\rm cool}$ to decrease in the inner parts, this time-scale increases at larger radii. Even though $Q$ is less than unity, since the dimensionless cooling time is long, the disc will not collapse complete but will be in a self-regulated self-gravitational state.
 Also, the effect of the self-gravity is more evident at high accretion rates. But for $\alpha_{0}=0.3$ the cooling time-scale is smaller than unity at all radii. Again, the self-gravity of the disc slightly  causes the cooling time-scale to decrease in the inner parts, and a longer time-scale in the  outer parts. So, when the Toomre parameter becomes less than unity, the disc becomes gravitationally unstable. In this case, the self-gravitating radius decreases because of the self-gravity of the disc, in particular at the high accretion rates.

Using equations (\ref{eq:QQ}) and (\ref{eq:beta1}) we can
calculate  the self-gravitating radius $R_{\rm sg}$ at which the
Toomre parameter becomes unity and the mass of the fragments $M_{\rm
frag}$ at this radius. Figure \ref{fig:figure6} shows results of the calculations for
a system with a central mass of $10^{8} M_{\odot}$, both for the solutions with MN prescription (top plots) and solutions with
free fraction $f$ (bottom plots). The other input parameters for the MN prescription are  $\mu=1.0$, $\mu_{\rm m}=0.6$, $\hat{k}=1$ and
$l_{\rm E}/\epsilon = 1$ and 10. For comparison, while solid curves  correspond to our solutions, the solutions neglecting self-gravity of the disc are shown by dashed curves. For the viscosity coefficient $\alpha_0$, we restrict to those values which  correspond to a nondimensional cooling time-scale lower than unity. For values of the coefficient $\alpha_0$ close to $0.5$, the  self-gravitating radius decreases.  But the location of this radius is more sensitive to the viscosity coefficient for low accretion rates either with self-gravity or without it. Interestingly, the self-gravity of the disc causes the self-gravitating radius to decrease, in particular for low accretion rates. But as the viscosity coefficient increases, the self-gravitating radii of the solutions with self-gravity and without self-gravity become closer. On the other hand, we note that in the  MN prescription the fraction of dissipated energy into corona is directly proportional to $\alpha_0$. Thus, as more energy is transported from the disc to the corona, the location of the self-gravitating radius becomes more independent of the self-gravity of the disc. Figure \ref{fig:figure6} (top; right plot) shows the mass of the fragments at the self-gravitating radius in the case of the MN prescription. The
mass of the fragments increases significantly when the accretion
rate is high and, generally, the mass of the
fragments increases because of the existence of a corona (KS). But the self-gravitating radius decreases due to the
existence of the corona either with low or high accretion rate.
Consequently, the presence of a corona makes the disc more
gravitationally unstable (KS). Also, the self-gravity of the disc significantly lead in reduction of the mass of the fragments for all values of the viscosity coefficient or accretion rate.

For comparison, the bottom plots of Figure \ref{fig:figure6} show the self-gravitating radius and the mass of the fragments for the solutions with free parameter $f$. Each curve is labeled by a corresponding fraction $f$. The other input parameters are  $\mu=1.0$, $\mu_{\rm m}=0.6$, $\hat{k}=1$ and
$l_{\rm E}/\epsilon = 1$. Again, solutions without self-gravity of the disc are shown by the dashed curves. For the effect of the self-gravity, in this case,  behavior of the variations of $R_{\rm sg}$ and $M_{\rm frag}$ with the input parameter are qualitatively similar to the solutions with MN prescription: reduction of the self-gravitating radius due to the self-gravity of the disc, but as more energy is transported from the disc to the corona, location of the self-gravitating radius becomes independent of the self-gravity of the disc. However, the mass of the fragments significantly decreases due to the self-gravity of the disc, irrespective of the input parameters, notably the parameter $f$.

\section{Discussion}

Generally and  for simplicity, the radial or the vertical components
of the gravitational force due to the self-gravity of the disc have
been neglected in most of the analytical studies of self-gravitating
accretion discs. However, some authors emphasize on the importance
of the gravitational force due to the self-gravity of the disc
(e.g., Paczynski 1978; Hur\'e 1998; Bertin  \& Lodato 1999). To our
knowledge, most of the previous studies of star formation in AGNs or
the Galactic center do not consider the components of the
gravitational force due to the self-gravity of the disc (e.g.,
Goodman \& Tan 2004; Nayakshin \& Cuadra 2005). Although KS studied
the possible effects of a corona on the gravitational stability of an
accretion disc, they also neglected the components of the
self-gravity of the disc itself. In this work, we presented sets of
analytical solutions for the steady-state structure of
self-gravitating accretion discs with  corona. We also considered
the vertical component of the gravitational force due to the
self-gravity of the disc. The energy exchange between the disc and
the corona has been considered either as a function of the physical
variables of the disc, or as a free parameter of the model. Although there are
possible mass and angular momentum exchanges between the disc and
the corona, we have neglected these effects just for simplicity in order
to present a self-consistent model for a disc-corona system. Having
the analytical solutions, we studied  properties of the solutions, more importantly  gravitational stability of the disc. For the
conditions of fragmentation, not only the Toomre parameter but the
cooling time-scale are considered in our analysis.

 We showed that because of the energy
exchange, the corona lead in a cooler  the disc for all
the input parameters. This reduction of the temperature due to existence
of the corona, decreases the thickness of the disc. Since there is
no mass exchange between the disc and corona, the surface density
increases and as a result, the vertical component of the
gravitational force due to the self-gravity of the disc becomes more
effective. When the Toomre parameter reaches to around unity, the
ratio $\zeta$ is around $4.8$ which means that the vertical
component of the gravitational force due to the self-gravity of the
disc is a few times larger than the vertical component of the
gravitational force of the central object at the self-gravitating
radius. Thus, this term may modify
properties of the solutions as our solutions show. For example,
self-gravity of the disc in a disc-corona system causes the
surface density to decrease at the inner parts of the disc, though the temperature
of the disc is unchanged when  the self-gravity of the disc is
considered.

Considering Toomre parameter and cooling time-scale conditions for
fragmentation, solutions with small viscosity coefficients are
stable to fragmentation because in the inner parts of the disc $Q\gg
1$ and when Toomre parameter reaches to unity, the cooling time-scale
becomes greater than one. This result is independent of the
accretion rate or other input parameters. But for large viscosity
coefficient the cooling-time scale is smaller than one for all
radii, and so, once the Toomre parameter becomes less than one
the disc may fragment. However, the cooling time-scale slightly  decreases at the inner parts of the disc because of the self-gravity of the disc and
for high accretion rates, this time-scale increases.

Also, self-gravitating radius and the mass of the fragments at
self-gravitating radius decrease because of the self-gravity of the
disc. However, change of the self-gravitating radius due to the
self-gravity is negligible when the amount of the energy which is
transported from the disc to the corona is increased. But reduction
of the mass of the fragments because of the self-gravity is
independent of the fraction $f$. These results are obtained either
with free $f$ or MN prescription for the fraction of energy
exchange. Although we estimate the mass of the fragments in an
accretion disc with corona around supermassive blackholes, we can
not determine whether these fragments are sufficiently long-lived to
be able to evolve in protostars over many orbital periods.

\section*{Acknowledgments}

We thank an anonymous referee for very useful suggestions which have significantly improved the paper.
FK acknowledges the support of an Ad Astra PhD Scholarship from University College Dublin. PD is
grateful to the Dublin Institute for Advanced Studies where part of this research was carried out. We thank Cliona Golden for reading the paper and her useful comments.

{}

\end{document}